\documentclass[aps,prb,twocolumn,superscriptaddress]{revtex4}
\usepackage{amsmath,amssymb}
\usepackage{graphicx}
\usepackage{bm}

\newcommand{\ceau}{Ce$_3$Au$_3$Sb$_4$}
\newcommand{\laau}{La$_3$Au$_3$Sb$_4$}

\begin{document}

\preprint{LA-UR-09-08002}

\title{Crystal electric field effects and quadrupole fluctuations in 
Ce$_3$Au$_3$Sb$_4$ detected by Sb NQR}


\author{S.-H. Baek}
\email[]{sbaek.fu@gmail.com}
\thanks{Current address: IFW-Dresden, Institute for Solid State Research, Dresden, Germany}
\affiliation{Los Alamos National Laboratory, Los Alamos, New Mexico 87545, USA}
\author{H. Sakai}
\affiliation{Los Alamos National Laboratory, Los Alamos, New Mexico 87545, USA}
\affiliation{Advanced Science Research Center, Japan Atomic Energy Agency,
Tokai, Ibaraki 319-1195, Japan}
\author{H. Lee}
\affiliation{Los Alamos National Laboratory, Los Alamos, New Mexico 87545, USA}
\author{Z. Fisk}
\affiliation{Los Alamos National Laboratory, Los Alamos, New Mexico 87545, USA}
\affiliation{Department of Physics \& Astronomy, University of California, 
Irvine, CA 92697, USA} 
\author{E. D. Bauer}
\affiliation{Los Alamos National Laboratory, Los Alamos, New Mexico 87545, USA}
\author{J. D. Thompson}
\affiliation{Los Alamos National Laboratory, Los Alamos, New Mexico 87545, USA}

\date{\today}

\begin{abstract}
We report $^{121,123}$Sb NQR studies on single crystals of the narrow gap 
semiconductor \ceau. 
Five NQR lines from the two Sb nuclei were successfully identified.  
The temperature 
dependence of the nuclear quadrupole frequency ($\nu_Q$), as well as the static
magnetic susceptibility ($\chi$), is well explained by crystal electric field effects. 
The nuclear spin-lattice relaxation rates ($T_1^{-1}$) 
of both $^{121}$Sb and $^{123}$Sb increase rapidly with decreasing 
temperature. The ratio of $T_1^{-1}$ for the two Sb isotopes is constant at high 
temperatures but it decreases at low 
temperatures, indicating the role of quadrupole fluctuations of the Ce ions.  
The possible origin of the large specific heat at low 
temperatures is discussed basing on our results.  

\end{abstract}



\maketitle

\section{Introduction}

Cerium- or uranium-based ternary compounds in the form of A$_3$T$_3$X$_4$ 
(``\textit{334}'', A = Ce,U; T = transition metal elements; X = Sb, Bi) exhibit unusual 
physical properties. When T=Ni,Pd,Pt, the compounds have a trend to become so-called 
Kondo insulators that feature a small gap originating from the hybridization between $f$ electrons and the conduction electrons.\cite{aeppli92} One of the most extensively studied 
members of these compounds is the Kondo insulator Ce$_3$Pt$_3$Bi$_4$,\cite{hundley90,reyes94} 
but there are  
also isostructural and isoelectric 
uranium-counterparts, U$_3$T$_3$Sb$_4$ (T=Ni,Pd,Pt)\cite{takabatake90,endstra90} and 
U$_3$Ni$_3$Bi$_4$.\cite{klimczuk08,baek09a}  

Ce$_3$Au$_3$Sb$_4$ also exhibits a narrow 
gap semiconductor-like behavior\cite{kasaya91} that  
looks very similar to Ce$_3$Pt$_3$Bi$_4$, so initially it was thought to be a
Kondo insulator.  However, the origin of the energy gap is  
different than in Ce$_3$Pt$_3$X$_4$  
(X=Sb,Bi), based on the facts that (1) La$_3$Au$_3$Sb$_4$ is an 
insulator\cite{takegahara92} unlike 
La$_3$Pt$_3$X$_4$ (X=Sb,Bi) that is metallic; and, (2) the Ce ion is well localized 
and trivalent,\cite{adroja95} but mixed valence is induced in Ce$_3$Au$_{3-x}$Pt$_x$Sb$_4$ 
and the concentration of Ce$^{4+}$ increases with increasing $x$.\cite{katoh96}  
Therefore, \ceau\ appears to be a unique ``\textit{334}'' 
compound in the sense that the $4f$ electrons are well localized, yet 
similar semiconducting properties with a   
narrow gap\cite{kimura93} with a different origin than in Kondo 
insulators. 

An anomalous behavior which is unsettled yet in \ceau\ is that its specific heat 
shows a broad peak centered at 1 K resulting in a logarithmic increase of $C/T$. 
The large electronic specific heat coefficient $\gamma$, which increases up to 
$\sim 4$ J/mol-K$^2$ at 0.5 K, 
cannot be understood in a semiconductor with very low density of charge carriers. 
Although the transport properties are very 
sensitive to sample quality, the large $\gamma$ itself at 
low temperature is commonly observed even in a high quality single 
crystal,\cite{lee07} indicating that the  
large $\gamma$ is an intrinsic property.  One scenario suggested for this 
observation is that a  
\textit{localized} narrow $f$-band is located at the Fermi level.\cite{lee07} Although 
this can explain the large $\gamma$ at low temperature and low carrier density 
simultaneously, it is not clear why the $f$ band which is 
assumed to lie exactly at the Fermi level should remain localized.   

Nuclear quadrupole resonance (NQR) is an ideal technique to resolve 
this ambiguity because it is a sensitive probe of local spin and charge
fluctuations.  In this paper, we report $^{121,123}$Sb NQR studies 
on  single crystals of \ceau.  The exact diagonalization of the crystal 
electric field Hamiltonian gives a $\Gamma_6$ doublet ground state, and this 
scheme provides excellent agreement with measurements of the uniform 
magnetic susceptibility, the magnetization, and the nuclear quadrupole frequency.
The nuclear spin-lattice relaxation rates of both $^{121}$Sb and $^{123}$Sb in 
the temperature range 2--300 K reveal a contribution from  
quadrupole fluctuations at low temperatures.  

\section{Sample preparation and experimental details}

Single crystals of \ceau\ and \laau\ were grown as described in ref.~\onlinecite{lee07}. 
In order to increase the \textit{rf} penetration depth and the filling factor,
the single crystals were ground into powder.
Though the overall crystal symmetry of \ceau\ is body-centered cubic (space 
group $I\overline{4}3d$), Ce and Sb atoms are located at positions that locally have 
tetragonal symmetry.

$^{121}$Sb ($I=5/2$) and $^{123}$Sb ($I=7/2$) NQR experiments were  
carried out in zero field using a conventional 
phase-coherent pulsed spectrometer in the range of temperature 2--300 K.  The Sb NQR 
spectra were obtained using the Hahn echo sequence i.e., $\pi/2-\tau-\pi$ with 
a typical $\pi/2$ pulse length $\sim 2\mu$s and a separation $\tau\sim 8\mu$s.     
The nuclear quadrupole frequency ($\nu_Q$)  
was determined by the separation between equally spaced transitions, as 
expected for axial local symmetry (anisotropy parameter $\eta=0$). 
The nuclear spin-lattice relaxation  
rate ($T_1^{-1}$) was measured by the saturation recovery at the $2\nu_Q$ transition 
for $^{123}$Sb 
and the $1\nu_Q$ transition for $^{121}$Sb. The relaxation data were then fitted to 
the appropriate equation for each transition.   

\section{Results and discussion}

\subsection{Crystal electric field Hamiltonian}

Due to localization of the $f$ moment of the Ce$^{3+}$ ion in \ceau\, one 
may expect that the crystal  
electric field (CEF) effect dominates the basic magnetic and electronic properties.  
Indeed, two well-defined CEF transitions at 25 K and 150 K were 
detected by inelastic neutron  
scattering (INS).\cite{adroja95} The CEF Hamiltonian in  tetragonal symmetry can be 
written as\cite{hutchings65} 
\begin{equation}
\label{eq:cef}
\mathcal{H}_\text{CEF} = B_2^0O_2^0+B_4^0O_4^0+B_4^4O_4^4,
\end{equation}
where $O_l^m$ are the Stevens operators\cite{stevens52} and the CEF parameters are 
$B_2^0=3.528$ K, $B_4^0=0.244$ K, and $B_4^4=2.077$ K.\cite{adroja95}  
By exact diagonalization of the CEF Hamiltonian, we obtained three doublet 
eigenstates with the ground state $\Gamma_6=|\pm 1/2\rangle$.  The first and second
excited states were identified to be  $-a|\pm 5/2\rangle + b|\pm 3/2\rangle$ and 
$\pm a|\pm 3/2\rangle \pm b|\mp5/2\rangle$, respectively, where $a=0.7561$ and 
$b=0.6844$. The schematic CEF level diagram is shown in the inset of Fig.~1. 
It should be noted that the  
CEF $\Gamma_6$ ground state carries the electric quadrupole moment as  
well as the magnetic dipole moment. 

\begin{figure}
\centering
\includegraphics[width=\linewidth]{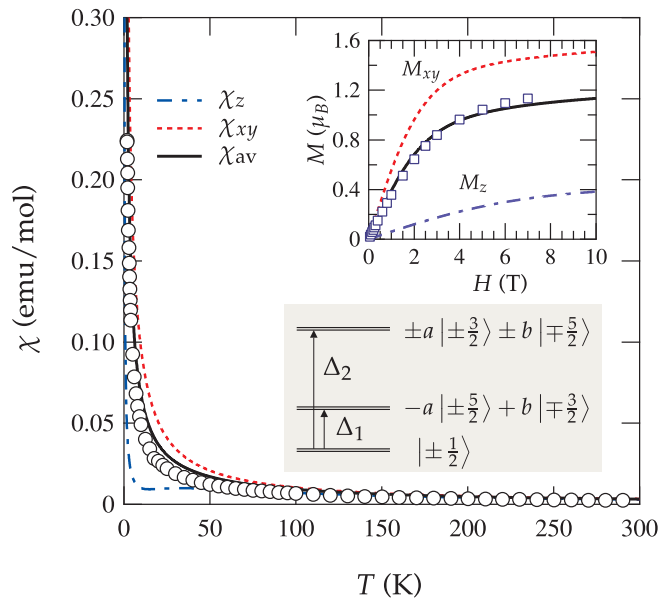}
\caption{\label{fig:chi}
Magnetic susceptibility $\chi$ measured at 0.1 T as a function of temperature.  
Magnetization $M$ at 2 K also is shown in the inset. Dotted lines are results from CEF 
calculations, and their powder averages are shown as thick solid lines.  The 
CEF level scheme   
is drawn with corresponding eigenstates, in which  
$a=0.7561$, $b=0.6844$, $\Delta_1\sim 25$ K, and $\Delta_2\sim 150$ K.   }
\end{figure}

The calculated susceptibility $\chi$ and magnetization $M$ from 
Eq.~(\ref{eq:cef}) agree with the experimental data as shown in 
Fig.~\ref{fig:chi}. Although the measurement was made in a single crystal, the 
powder averages fit the   
data satisfactorily, because there are three local principal axes along [111] 
at the Ce sites.  

\subsection{Temperature dependence of nuclear quadrupole frequency, $\nu_Q$}

\begin{figure}
\centering
\includegraphics[width=\linewidth]{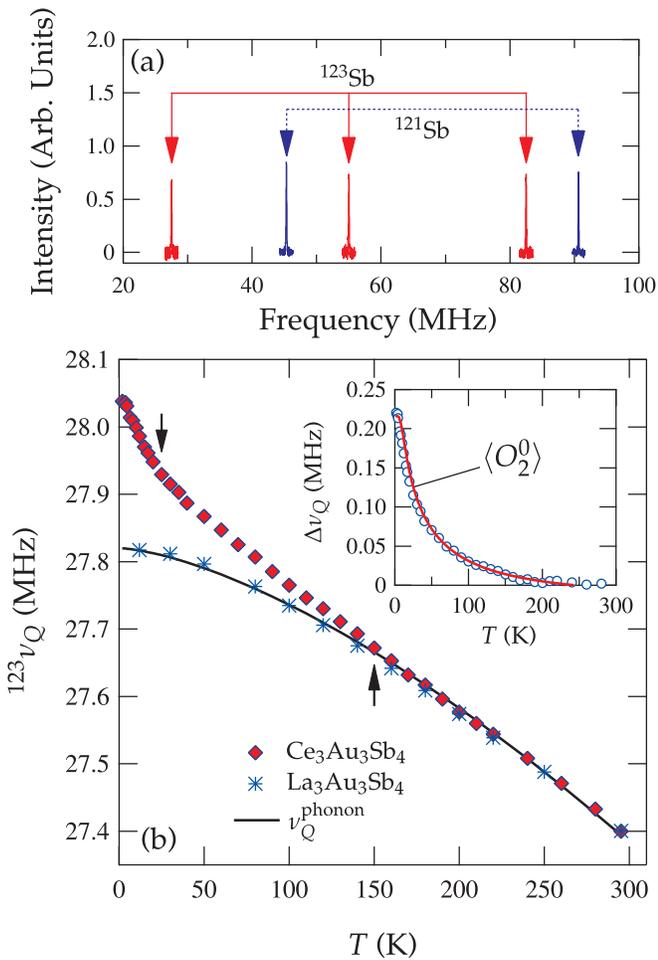}
\caption{\label{fig:nuQ}
(a) Sb-NQR spectra obtained at $T=240$ K in zero field. Spectra of both Sb 
isotopes are equally spaced  
as expected in uniaxial symmetry. At this temperature, $\nu_Q$ is 27.50 MHz and 45.32 MHz for 
$^{121}$Sb and $^{123}$Sb, respectively.   (b) $\nu_Q$ of $^{123}$Sb as a 
function of temperature. Results for \laau\, which were scaled down by $\sim$2\% 
for comparison, are shown together with the expected phonon contribution 
$\nu_Q^\text{phonon}$. Inset: Difference between \ceau\ data and 
$\nu_Q^\text{phonon}$. This difference  
is well explained by the expectation value of the quadrupole moment 
$|\langle O_2^0\rangle|$ which was scaled to data. 
}
\end{figure}

The NQR spectra of both $^{121}$Sb and $^{123}$Sb are shown in 
Fig.~\ref{fig:nuQ} (a). A total of five lines from the two Sb isotopes is found and,
as expected from the local uniaxial threefold symmetry at the Sb 
sites, the lines for each of the Sb nuclei are equally spaced and the
nuclear quadrupole frequency, $\nu_Q=3e^2qQ/2h(2I-1)$, correctly scales with the nuclear 
quadrupole moment $Q$ and the nuclear spin $I$ for the two Sb isotopes. For  
$Q$, we used the values from ref.~\onlinecite{tou05} 
($^{121}Q=-0.597\times10^{-28}$  m$^{2}$ and $^{123}Q=-0.762\times10^{-28}$ m$^{2}$).

We measured the temperature dependence of $\nu_Q$ for $^{123}$Sb, tracking 
the $2\nu_Q$ transition, which is shown in Fig.~\ref{fig:nuQ} (b). 
The strong temperature dependence of $\nu_Q$ indicates that 
the electric field gradient (EFG) $eq$ increases with decreasing temperature. 
There should be a thermal (phonon) contribution that is phenomenologically 
described\cite{kaufmann79} by 
the relation of $\nu_Q^\text{phonon}=\nu_Q^0-a T^{3/2}$ that 
is drawn as a solid line in Fig.~\ref{fig:nuQ} (b). Clearly, there is an 
additional contribution below $\sim 150$ K.  For comparison, we measured 
$^{123}\nu_Q$ of \laau.  Because the the temperature dependence of $\nu_Q$ for \laau\ is 
explained solely by the phonon contribution, 
the additional increase of $\nu_Q$ for \ceau\ at low temperatures should 
arise from the Ce ions. We find that, as denoted by 
arrows in Fig.~\ref{fig:nuQ} (b), features in  
$\nu_Q$ occur at the  
CEF splitting energy $\Delta_1\sim 25$ K and $\Delta_2\sim 150$ K, suggesting 
that $\nu_Q$ is affected by CEF splitting.   This is  
clearly shown in the inset in which  
$\Delta\nu_Q=\nu_Q(\text{Ce})-\nu_Q(\text{La})$ is proportional to the thermal 
average of the expectation value of the quadrupole  
moment,
\begin{equation}
\label{eq:Qmoment}
\langle O_2^0 \rangle (T)= \sum_m \frac{\langle O_2^0\rangle_m \exp(-E_m/T)}{Z},
\end{equation}
where $E_m$ and $\langle O_2^0\rangle_m$ are the energy and the quadrupole 
moment, respectively, of $m$th sublevel of $J=5/2$ and $Z$ is the 
partition function.  $\langle O_2^0\rangle (T)$ increases rapidly with decreasing 
temperature due to the uniaxially 
elongated charge distribution in the $\Gamma_6$ ground state.  
A similar enhancement of $\nu_Q$ was observed in an Sb NQR study of the 
skutterudite compound PrOs$_4$Sb$_{12}$, and the origin  
was explained by the coupling of the Sb nuclear quadrupole moment with the 
hexadecapole moment of Pr$^{3+}$ ($f^2$) which can have a finite thermal average,  
unlike other multipole moments.\cite{tou05}
Since the hexadecapole moment 
that requires a $\Gamma_1$ state is not allowed in \ceau, 
the increase of $\Delta\nu_Q$ should result from the direct coupling between 
the quadrupole moment of   
the Ce ions and the Sb nuclear quadrupole moment. We emphasize here that this 
direct correlation of $\nu_Q$ with the crystal field splittings and the  
quadrupole moments of an ion is extremely rare, if any, in Ce-based compounds. 

\subsection{Nuclear spin-lattice relaxation rate and quadrupole fluctuations}

\begin{figure}
\centering
\includegraphics[width=\linewidth]{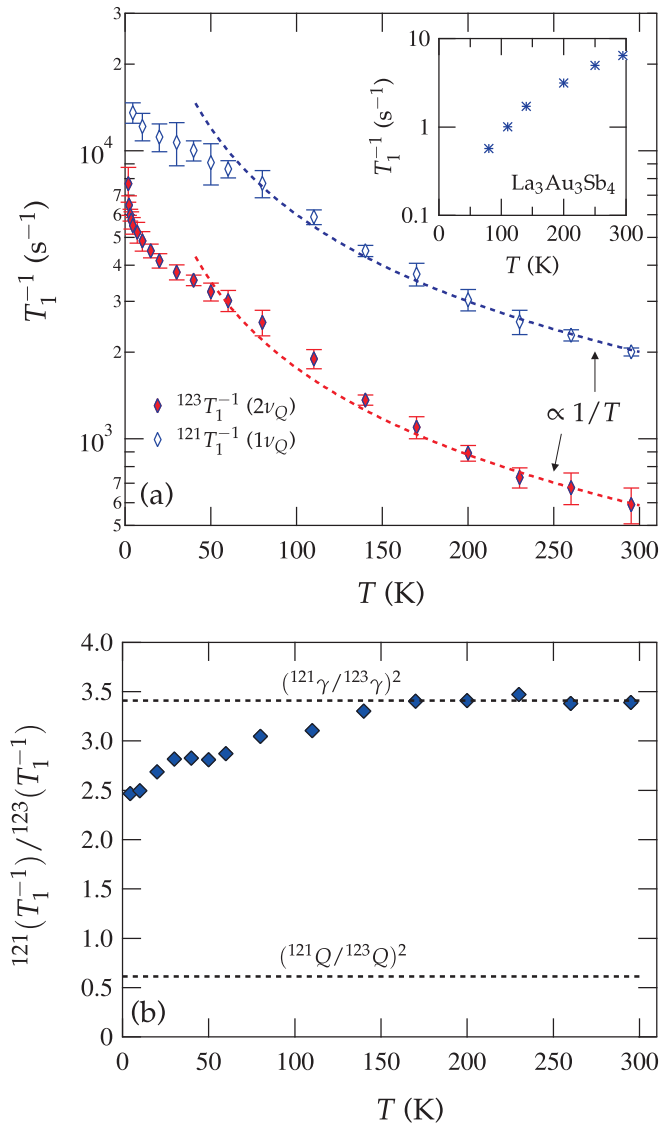}
\caption{\label{fig:T1}
(a) $T_1^{-1}$ as a function of temperature for both $^{121}$Sb and $^{123}$Sb. 
Both Sb nuclei  
show the same $1/T$ dependence in the high temperature region.  
At low temperatures, below $\Delta_2\sim 150$ K, the temperature dependence of 
$^{123}T_1^{-1}$ becomes   
distinguished from that of $^{121}T_1^{-1}$. INSET: $^{123}T_1^{-1}$ measured 
at $2\nu_Q$ transition in \laau\ decreases rapidly with lowering 
temperature.  
(b) Temperature-dependent ratio between $^{121}T_1^{-1}$ and $^{123}T_1^{-1}$. 
This ratio deviates from the value  
expected for magnetic fluctuations, revealing a significant contribution from 
quadrupole fluctuations at low temperatures.} 
\end{figure}

So far, we have shown that the static properties $\chi$, $M$, and $\nu_Q$ are 
dominated by CEF effects.  To study the local dynamics, we have measured 
$T_1^{-1}$ for both $^{121}$Sb and $^{123}$Sb (Fig.~\ref{fig:T1}).   
We find that $T_1^{-1}$ increases rapidly with decreasing temperature for both 
$^{121}$Sb and $^{123}$Sb.  In 
contrast, for \laau\,  
$^{123}T_1^{-1}$, which measured at $2\nu_Q$ transition, decreases 
exponentially with decreasing temperature as shown in the  
inset of Fig.~\ref{fig:T1}(a), which indicates that \laau\ is a simple band-gap insulator.  
Therefore, the fast $T_1$ in \ceau\ also arises from the Ce ions. 

In high temperature region, $T_1^{-1}$ for 
both Sb nuclei follows a $1/T$ dependence drawn as dotted lines in Fig.~\ref{fig:T1}.  
The origin of this $1/T$ dependence is not clear, but it could be ascribed to hybridization 
between the conduction electrons and the localized $f$ electrons in the 
semi-metallic region at high temperatures. Alternately, if $T_1^{-1}$ 
is dominated by magnetic fluctuations in a system of localized $f$ electrons, 
$T_1^{-1}$ is proportional to $A_\text{hf}^2\mu_\text{eff}/J_\text{ex}$ where 
$A_\text{hf}$ is the hyperfine coupling constant, $\mu_\text{eff}$ the 
effective magnetic moment, and $J_\text{ex}$ the  
exchange coupling.\cite{moriya56a} In this case, a decrease of $J_\text{ex}$ 
with decreasing $T$ may be responsible for the $T$ dependence of $T_1^{-1}$.  

In the temperature region below $\sim 150$ K, the temperature dependence of $T_1^{-1}$ 
for the two Sb isotopes become different, which is more prominent below $\sim 25$ K.  
In general, $T_1^{-1}$ due to spin 
fluctuations is written as\cite{moriya63} 
\begin{equation}
\label{eq:T1}
T_1^{-1} \propto T \gamma_n^2 \sum_q A_\text{hf}^2(q)\chi''(q,\omega_0),
\end{equation}
where $\gamma_n$ is the nuclear gyromagnetic ratio, $A_\text{hf}(q)$ the $q$-dependent 
hyperfine coupling constant, and $\chi''$ the imaginary part of the  
dynamic susceptibility. Since the ratio of $T_1^{-1}$ for the two Sb isotopes 
is the same as $(^{121}\gamma_n/^{123}\gamma_n)^2=3.41$, as shown 
in Fig.~\ref{fig:T1} (b), we conclude that  
magnetic fluctuations from Ce ions are the major contribution to relaxation 
processes at high temperatures, 
regardless of the origin of $1/T$ behavior. 

The ratio $T_1^{-1}$ for the two isotopes deviates from 3.41 
below $\Delta_2=150$ K toward the value expected for the ratio of the square 
of quadrupole moments  
$(^{121}Q/^{123}Q)^2$. This observation appears to be consistent with the 
enhancement of $\nu_Q$ at low temperatures.  
This trend is even more prominent below $\Delta_1$.  
In this temperature region, $^{123}T_1^{-1}$ shows an almost diverging behavior, 
while $^{121}T_1^{-1}$ increases more weakly down to 4 K. Unfortunately, $T_1$ 
is too short to measure below 4 K and 2 K, for $^{121}$Sb and $^{123}$Sb, 
respectively. Further, due to the shortening of $T_2$, the signal becomes very weak. 
Nevertheless, the different temperature dependences of the two isotopes 
indicate the important role of quadrupole fluctuations in the spin dynamics at low 
temperatures. However, this apparent contribution of the quadrupole 
fluctuations to the nuclear relaxation is incompatible with the fact that 
$\Gamma_6=\left|\pm1/2\right>$  
state alone cannot produce the  
quadrupole fluctuations due to the Kramers degeneracy of the ground state.\cite{abragam70} 
Here, we conjecture a finite matrix element between $\Gamma_6$ 
and the first excited doublet forming an effective ``quartet'' state. In this case, broadening 
of the levels, for example, due to anharmonic phonons\cite{thalmeier84} may help promote the 
mixture of the two doublet  
states. We recall a similar situation in a Yb-monopnictide 
YbSb whose ground state is also a $\Gamma_6$ doublet that consist of 
$\left|\pm 1/2\right>$ and $\left|\mp 7/2\right>$. YbSb shows a phase 
transition at 5 K and quadrupole order has been suggested as its origin, 
even with a Kramers ground doublet, which has no  
quadrupole moment, separated by rather large splitting energy of 170 K with the first 
excited states.\cite{hashi01,oyamada04}  
In our case, in comparison with YbSb, a mixed quadrupolar interaction seems to 
be more realistic, since the $\Gamma_6$ in \ceau\ has a 
quadrupole moment and the splitting energy to the first  
excited states is much smaller than that 
of YbSb.

The almost diverging $T_1^{-1}$ of $^{123}$Sb resembles critical 
slowing down near an ordering temperature and, indeed,  
specific heat measurements find a clear anomaly at 0.2 K.
From magnetic field dependence of the specific 
heat, the anomaly appears to be 
antiferromagnetic in nature.\cite{kurita} Although AFM order appears to be primary, 
the additional quadrupole contribution may be detected by the specific heat, and the 
coupled magnetic and quadrupolar interactions may lead to 
the observed large $C/T$ at low temperatures.  
In order to pin down the nature of the transition at 0.2 K, magnetization measurements 
at low temperatures are needed. Regardless of the detailed nature of the transition, the
temperature dependence of both $T_1^{-1}$ and $\nu_Q$ imply a strong coupling between the  
quadrupole moment of the Sb and the quadrupole moment of Ce in the CEF scheme.

\section{Conclusion}

In conclusion, we have measured Sb NQR in the narrow gap semiconductor \ceau\ and  
have shown that the static ($\nu_Q$) and dynamic 
($T_1^{-1}$) properties of \ceau\ are affected strongly by the crystal electric 
field splitting. The nuclear quadrupole frequency $\nu_Q$, after subtracting 
the phonon contribution, increases  
with deceasing temperature in proportion to the expectation value of  
the quadrupole moment of the Ce ion.  This 
reveals a clear coupling between the quadrupole moment of Ce and the 
nuclear quadrupole moment at Sb.  Furthermore, 
quadrupole fluctuations contribute to $T_1^{-1}$ at low 
temperatures in addition to magnetic fluctuations that 
are dominant. This finding strongly suggests the finite matrix element between the $\Gamma_6$ 
ground doublet and the excited doublet through intersite quadrupole-quadrupole 
interactions since the Kramers degeneracy does not  
allow the quadrupole fluctuations.  
From our NQR results, we propose that the 
anomalous divergence of $C/T$ at low temperatures may arise from the
quadrupole fluctuations. 
 
We have shown that \ceau\ is a special member of the 
\textit{334} family,   
not only due to its well localized $f$ moment with narrow-gap, semiconductor-like behavior
but also to the strong influences of crystal electric fields on 
the static and the dynamic local properties. It deserves further detailed experimental 
and theoretical studies to elucidate the role of quadrupole moments and 
their fluctuations in determining physical properties. 

\section*{Acknowledgement}

We thank V. Kataev, S. Kambe, and Y. Tokunaga for the useful suggestions and discussions.  
Work at Los Alamos National Laboratory was performed under the
auspices of the US Department of Energy, Office of Science.

\bibliography{mybib}

\end{document}